\documentclass[12pt]{article}
\usepackage[T1]{fontenc}
\usepackage[utf8]{inputenc}

\usepackage[top=4cm, bottom=4cm, left=3cm, right=2.5cm]{geometry}

\usepackage{setspace}
\usepackage{footmisc}
\usepackage[hidelinks]{hyperref}
\usepackage[round]{natbib}


\usepackage{turnstile}
\usepackage{cancel}
\usepackage{amssymb}
\renewcommand{\restriction}{\mathord{\upharpoonright}}

\usepackage{mathtools}
\usepackage{stackengine}
\setstackEOL{\\}
\usepackage{tikz}
\usepackage{tikz-3dplot}
\usepackage{subcaption}
\usepackage{tikz-cd}
\usetikzlibrary{arrows,decorations.pathmorphing,backgrounds,positioning,fit,petri,calc,shapes.misc,decorations.markings,intersections}

\usepackage{amsmath,amsfonts,amsthm} % Math packages
\usepackage{stmaryrd}
\usepackage{enumitem}
\usepackage{caption}
\usepackage{array}

\begin{document}

\title{Betting on Quantum Objects}
\author{Jeremy Steeger}
\date{\today}
\maketitle

\begin{abstract}
Dutch book arguments have been applied to beliefs about the outcomes of measurements of quantum systems, but not to beliefs about quantum objects prior to measurement. In this paper, we prove a quantum version of the probabilists' Dutch book theorem that applies to both sorts of beliefs: roughly, if ideal beliefs are given by vector states, all and only Born-rule probabilities avoid Dutch books. This theorem and associated results have implications for operational and realist interpretations of the logic of a Hilbert lattice. In the latter case, we show that the defenders of the eigenstate-value orthodoxy face a trilemma. Those who favor vague properties avoid the trilemma, admitting all and only those beliefs about quantum objects that avoid Dutch books.
\end{abstract}

\tableofcontents

\newpage

%%%%%%%%%%%%%%%%%%
%% Introduction %%
%%%%%%%%%%%%%%%%%%

\section{Introduction}\label{sec:intro}
% \noindent
% \textbf{1. Introduction.}\label{sec:intro} 
Judging the rationality of beliefs about the quantum world is a hard problem. To address it, it seems it might be useful to appeal to a general principle long thought to serve as a necessary condition for \emph{any} sort of belief to be rational: the avoidance of Dutch books. Bayesians often use the avoidance of Dutch books, or betting strategies that cause an agent to lose money no matter what the world is like, as a litmus test for the rationality of belief. This test already has a breadth of applications. Recent works by \cite{Paris2001} and \cite{Mundici2006}, to name but a few, have applied a generalized version of the classic synchronic Dutch book argument to beliefs in the propositions of various nonclassical logics. Not among these logics are those of Hilbert lattices (these will be detailed in Section 3). The formal language of Hilbert-lattice logics may be used to straightforwardly describe either of two sorts of quantum phenomena. Propositions in this language may refer to the outcomes of measurements of quantum systems, on the one hand, or to the properties of quantum objects underlying these outcomes, on the other.\footnote{By ``quantum objects'' we mean, quite simply, the referents of expressions such as ``electron'' or ``proton''---particles whose properties are often taken to be represented by bounded, self-adjoint operators on a Hilbert space. Our notion of \emph{object} is quite thin. We do not impose objects satisfy Leibniz's Law, or any variant of it. For work addressing questions of discernibility in this context, see Saunders's argument (\citeyear{Saunders2006}) and Hawley's response (\citeyear{Hawley2006}).} It is the project of this paper to apply the generalized Dutch book argument to the logics of Hilbert lattices. In accordance with the two ways these logics may be used, this application provides a means for assessing both beliefs about outcomes \emph{and} beliefs about objects.

\begin{figure}
    \centering
    \begin{tikzpicture}
        \newdimen\R
        \newdimen\nodesize
        \R=2 cm
        \nodesize=2 pt
        
        % corners
        \foreach \x/\l/\p in
         {60/a/above,120/b/above, 180/c/left,240/d/below,300/e/below,360/f/right}
         \coordinate [label=\p: ] (\l) at (\x:\R);
        
        % sides
        \foreach \x/\l/\p in {a/b/ab,b/c/bc, c/d/cd,d/e/de,e/f/ef,f/a/fa}
        \path[name path global/.expanded=\p] (\x)--(\l);
        
        % intersection paths
        \foreach \x/\l in
         {80/abde1,100/abde2, 140/bcef1,160/bcef2,200/cdfa1,220/cdfa2}
         \path[name path global/.expanded=\l] (\x:-\R)--(\x:\R);
        
        % intersections
        \foreach \x/\l/\p in
         {ab/abde1/ab1,ab/abde2/ab2,
         de/abde1/de1,de/abde2/de2,
         bc/bcef1/bc1,bc/bcef2/bc2,
         ef/bcef1/ef1,ef/bcef2/ef2,
         cd/cdfa1/cd1,cd/cdfa2/cd2,
         fa/cdfa1/fa1,fa/cdfa2/fa2} \path [name intersections/.expanded = {of = {\x} and \l, by = \p}];
        
        % nodes
        \foreach \x in
         {a, b, c, d, e, f, ab1, ab2, bc1, bc2, cd1, cd2, de1, de2, ef1, ef2, fa1, fa2}
         \draw (\x) circle (\nodesize);
         
         % contexts
         \node[draw,fit=(a) (b)] {};
         \node[draw,rotate fit=-30, fit=(b) (c)] {};
         \node[draw,rotate fit=30, fit=(c) (d)] {};
         \node[draw,rotate fit=0, fit=(d) (e)] {};
         \node[draw,rotate fit=-30, fit=(e) (f)] {};
         \node[draw,rotate fit=30, fit=(f) (a)] {};
         \node[draw,rotate fit=90, fit=(bc2) (cd1) (ef2) (fa1)] {};
         \node[draw,rotate fit=-30, fit=(ab1) (fa2) (cd2) (de1)] {};
         \node[thick,draw,rotate fit=30, fit=(ab2) (bc1) (de2) (ef1)] {};
         
         % coloring
         \draw[black,fill=black] (e) circle (\nodesize);
         \draw[black,fill=black] (fa2) circle (\nodesize);
         \draw[black,fill=black] (b) circle (\nodesize);
         \draw[black,fill=black] (cd1) circle (\nodesize);
        
    \end{tikzpicture}
    \caption{Cabello and collaborators' 18-vector proof of the Kochen-Specker result. A white coloring for a node represents an assignment of 0 (impossible); a black coloring represents an assignment of 1 (possible). No coloring will yield precisely one black node for each box; one such failed coloring is shown.}
    \label{fig:Cabello}
\end{figure}
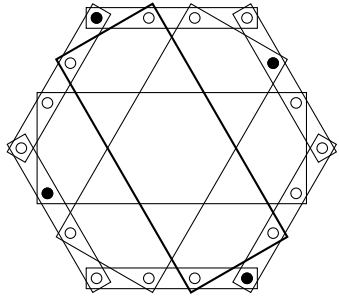

Dutch book arguments have thus far been applied to the former sort of beliefs but not to the latter.\footnote{As far as this author can tell, in any case.} The absence of such arguments is understandable---it is not immediately clear what quantum objects might look like, in the first place! Much of the difficulty in picturing quantum objects stems from the Kochen-Specker theorem. Kochen and Specker proved fifty years ago that there can be no hidden variable theory underlying the possible outcomes of measurements of quantum systems that both uniquely determines these outcomes and renders them independent of the context of measurement (\citeyear{Kochen1967}).\footnote{In this paper, we use ``contextuality'' to refer exclusively to dependence on the context of measurement, and not on the context of preparation or transformation; these latter sorts of dependence are explored by Spekkens (\citeyear{Spekkens2005}).} In particular, there can be no assignment of sharp and context-independent values for properties such that these values are faithfully revealed by measurement. Consider the simple proof of these results offered by Cabello and collaborators, illustrated in Figure \ref{fig:Cabello} (\citeyear{Cabello1996}). A given node corresponds to possible values that may be observed for certain spin properties of two entangled electrons. Each box stands in for a measurement context, a maximal set measurements that can be simultaneously performed. Whenever a box is measured, precisely one node in the box will assign the values observed. If we take measurement to reveal property-values that do not depend on context, then no consistent global assignment of values can be given.

The shock brought on by the impossibility of such an assignment is aptly reflected in subsequent approaches to both quantum logic and Bayesian methods in quantum foundations. A full and just review of the literature is well beyond the scope of this paper (let alone its introduction), but it suffices to point to a few highlights. \cite{Randall1976}, for instance, present a strongly operational Bayesian approach. They prove that Born-rule beliefs about outcomes of measurements satisfy a no-Dutch-book condition (we will return to this result in Section 5). The authors use symbols that refer to \emph{physical operations} (``well-defined physically realizable, reproducible procedure[s]'') and  \emph{outcomes}; there is never any mention of the properties of unseen objects prior to measurement (\citeyear[p. 170]{Randall1976}). Randall and Foulis's disavowal of such properties, if not their operationalism per se, was echoed decades later in Pitowsky's work on ``quantum gambles''---strategies for betting on the outcomes of incompatible measurements, only one of which is to be performed by a bookie. In this work, the outcomes of measurements are treated as mere possibilities ``[not associated] with properties that exist prior to the act of measurement'' (\citeyear[p. 408]{Pitowsky2003}).

Meanwhile, logicians set about developing realist semantics for Hilbert-lattice logics that \emph{do} refer to such properties. Putnam's proposal is perhaps the most infamous. On this proposal, for every box in Figure \ref{fig:Cabello}, the disjunction of all the property-ascriptions contained in that box is true and only one of the disjuncts is true. Putnam appears to maintain that the truth of these disjuncts does not depend on measurement context, asserting just that (since we can only measure one box at a time) we may be ignorant as to which disjunct is the true one (\citeyear[p. 186]{Putnam1975}). Quite clearly, this prescription runs afoul of the Kochen-Specker setup in the figure. No such truth valuation exists!%\footnote{Demopoulos salvages a realist semantics quite similar to Putnam's by letting truth valuations depend on measurement context \cite{Demopoulos1976}. We will not consider such contextual semantics in this paper.}

But the failure of Putnam's proposal does not demonstrate that no noncontextual, realist semantics for quantum logics are forthcoming. There are, after all, extant proposals for interpreting quantum mechanics as a theory about the properties of observer-independent objects that do not run afoul of the Kochen-Specker result. The eigenstate-value link is perhaps the oldest. More recently, a proposal using vague properties has been developed (\citealt{Lewis2016,Pykacz2015}). Each of these proposals may be perspicuously described as truth-value semantics for Hilbert-lattice logics (this exercise will be conducted explicitly in Section 6).

Pitowsky suggests that Putnam's approach comes with ``a heavy price-tag: the repudiation of classical [propositional] logic'' (\citeyear[p. 408]{Pitowsky2003}). Insofar as we continue to work with Hilbert-lattice logics, the proposals sketched here follow Putnam in this ``repudiation.'' But it is far from clear that classical propositional logic is threatened by its failure to describe an ontology of quantum objects with certain properties. One may well think that this familiar logic has a vast domain of applicability in metaphysics, which applicability is not particularly troubled by a nonclassical attitude towards the quantum. The supposed repudiation, then, may be quite inexpensive.

Given this plausible lack of expense---and given the utility of object-talk in homespun metaphysical practice---it seems natural to wonder about the rationality of the beliefs the sketched proposals suggest we have about the properties of quantum objects prior to measurement. Certainly, these properties are inaccessible to us. We let go of the idea that a bookie can always be some human able to look at the results on which bets hinge, and consider the bookie as some hypothetical entity who always has perfect knowledge of the quantum system. That done, we are free to use the generalized synchronic Dutch book argument as an in-principle rationality constraint to assess our beliefs about quantum objects in addition to our beliefs about the outcomes of measurements of them.

We apply this constraint by proving a quantum probabilism theorem. Treating the projection lattice $\mathcal{P(H)}$ of a finite-dimensional Hilbert space $\mathcal{H}$ as a quantum logic, if the possible ideal beliefs an agent should have regarding propositions in $\mathcal{P(H)}$ are given by the restrictions of vector states to the lattice, then the Born-rule probabilities are all and only the total belief functions avoiding Dutch books. This theorem extends previous results due to K\"{u}hr and Mundici, described in Section 2, in a novel direction (\citeyear{Kuhr2007}). Section 3 provides a brief primer on the Hilbert-lattice formalism used for our quantum logics, and Section 4 presents the proof of the main theorem.

We proceed in Section 5 to demonstrate how an analogue of the operationalist Dutch book result of Randall and Foulis can be recovered in our framework. In Section 6, we complete the translation of our realist proposals into semantics for quantum logics, and we use our main theorem to assess vague-property semantics.

Finally, in Section 7, we show that the defenders of the eigenstate-value orthodoxy face a trilemma. They must choose one of: not using Born's rule to fix agents' beliefs; suggesting agents have no degree of belief in many property-ascriptions; or leaving agents susceptible to Dutch books. Those who favor vague properties avoid this trilemma, admitting all and only those beliefs about quantum objects that avoid Dutch books. Section 8 suggests directions for future work. Proofs of all results can be found in the appendix.

%%%%%%%%%%%%%%%%%%%%%%%%
%%%%%%%%%%%%%%%%%%%%%%%%
\section{Generalizing Dutch-bookability for quantum logics}\label{sec:gen}

% \vspace{\baselineskip}
% \noindent
% \textbf{2. Generalizing Dutch-Bookability for Quantum Logics.}\label{sec:gen}
The heart of the generalized synchronic Dutch book argument is a game played by an agent and a bookie, a betting arrangement testing the agent's beliefs about the state of the world. The state of the world is some assignment of truth values to a set of propositions, $L$. %We take $L$ to be a propositional language: it contains propositional constants $\phi$, distinguished elements $\top$ and $\bot$, and it is closed under the negation operation, $\neg$, and the and countable meet and join operations, $\bigwedge_{i\in I}$ and $\bigvee_{i\in I}$.\footnote{We work with infinitary languages, as this proves to be the most natural setting for our quantum logics.}
Not all propositions may be assigned truth values, so we take a state of affairs to be a partial function $T:L\to N$, where $N$ is a set of truth values. $\mathbb{T}$ will denote the set of states of affairs; these are the subjects of the agent's beliefs and the bookie's knowledge.

We focus on a certain sort of doxastic attitude an agent may have regarding a proposition $\phi\in L$, namely the agent's \emph{degree of belief} or degree of confidence in $\phi$. Following standard Bayesian approach, we assume that if an agent has a degree of belief in $\phi$, then this degree may be represented by some value in the unit interval (\citealt{Ramsey1931a,Smith2014}). We say the agent's degrees of belief are given by some partial function $B:L\to [0,1]$; the function is partial to allow for cases where an agent withholds belief. Note there is a question of how an agent should fix her degree of confidence in a proposition that she suspects is neither true nor false. We return to this issue in Sections 5--7.

The bookie, for her part, assigns stakes $s_i\in\mathbb{R}$ to each proposition $\phi_i\in L$, and is presumed to have perfect knowledge of the state of the world. Given this knowledge, the bookie then decides what the agent's ideal degree of belief in a proposition should be. In the classical case, this decision is straightforward: if a proposition is true, the agent should have degree of belief one in it; if it is false, the agent should have degree of belief zero in it. But propositions may be ascribed esoteric truth values in non-bivalent logics, and so the decision is not always as clear. We thus make the decision explicit by assigning to each state of affairs $T$ a partial function $V_T :  L\to [0,1]$ describing what the agent's ideal beliefs should be in this case. Following Bradley, we call this function a \emph{cognitive evaluation} (\citeyear{Bradley2016}). $\mathbb{V}$ will denote the set of all such evaluations.

If the bookie decides the agent should not have any degree of belief in propositions in some subset of $L$, then $V_T$ will be partial. In other words, the bookie will not agree to betting arrangements concerning propositions for which $V_T$ is undefined. To identify subsets of $L$ on which she will accept bets, we stipulate that a set of propositions $\Gamma \subset L$ is \emph{testable} if there is some $V_T \in\mathbb{V}$ such that $\Gamma$ is included in the domain of $V_T$.

Now suppose our bookie sets some nonnegative stake $s_i \geq 0$ for the proposition $\phi_i$. Given our agent has degree of belief $B(\phi_i)$ in $\phi_i$, we suppose that the agent is willing to pay the bookie $B(\phi_i) \cdot s_i $ for the $s_i$-dollar bet on $\phi_i$. The bookie will pay out $s_i \cdot V_T(\phi_i)$ assuming $T$ describes the state of the world. We also allow ``reverse bets.'' For negative stakes $s_i < 0$, if the agent agrees to the bet, then the bookie pays the agent $B(\phi_i)\cdot |s_i | $, and the agent must pay $V_T(\phi_i)\cdot |s_i |$ when the bookie comes to collect. Now suppose there is some finite, testable set of propositions $\Gamma=\{\phi_1, \mathellipsis, \phi_n \}$ and an ordered set of stakes the bookie can give them,  $\langle s_1, \mathellipsis, s_n \rangle \in \mathbb{R}^n$, such that for all $V\in \mathbb{V}$ whose domain includes $\Gamma$,
\begin{equation}\label{eq:dutch_gen}
\sum_{i=1}^n s_i(V(\phi_i)-B(\phi_i)) < 0;
\end{equation} 
the agent will always lose money to the bookie when she places her bets on the propositions in $\Gamma$. Thus, we say that if there are sets $\Gamma$ and $\langle s_1, \mathellipsis, s_n \rangle$ satisfying equation (\ref{eq:dutch_gen}) for a belief function $B$ given cognitive evaluations $\mathbb{V}$, then $B$ is Dutch-bookable, or fails to avoid Dutch books.\footnote{Paris demonstrates that this condition generalizes that used in de Finetti's original synchronic Dutch book argument (\citealt{Paris2001,deFinetti1990}).}

This betting game suggests a tight link between coherent beliefs and the convex hull of the cognitive evaluations,  $\mathrm{co} (\mathbb{V} )$. This is the set of all finite convex sums of such evaluations, finite linear sums of $V\in \mathbb{V}$ where the coefficients are nonnegative and sum to one. \cite{Kuhr2007} demonstrate that so long as $\mathbb{V}$ contains only total functions and is closed under pointwise convergence, a total belief function $B$ avoids Dutch books if and only if it lives in the pointwise-closure of the convex hull of $\mathbb{V}$, $\overline{\mathrm{co}}(\mathbb{V})$.\footnote{Recall a set $F$ of functions $f:X\to Y$ is closed under pointwise convergence if, for any net $\{f_i\}$ in $F$ that converges pointwise to $f$, $f$ is in $F$; a net $\{f_i\}$ converges pointwise to $f$ if, for all $x\in X$ and all $\epsilon >0$, there is some $i'$ such that for all $i\geq i'$, $|f_i(x)-f(x)|<\epsilon$. The closure $\overline{\mathrm{co}}(\mathbb{V})$ is just the smallest pointwise-closed set containing  $\mathrm{co}(\mathbb{V})$.}
\begin{quote}\label{thm:Mundici}
\textbf{Theorem 1.} [Theorem 2.3 in \cite{Kuhr2007}] For $\mathbb{V}$ pointwise-closed in $[0,1]^L$, a belief function $B\in [0,1]^{L}$ does not admit a Dutch book if and only if $B \in \overline{\mathrm{co}}(\mathbb{V})$.
\end{quote}
\noindent
We include a proof of this result in the appendix in order to demonstrate the proceeding corollaries.

% For our purposes, we prove a modest generalization of K\"{u}hr and Mundici's theorem---$\mathbb{V}$ need not be pointwise-closed for the theorem to hold, so long as all sets of restrictions of the functions in $\mathbb{V}$ to a finite set are pointwise-closed. This proves to be useful when considering logics with a finite set $N$ of truth values; indeed, we will use it when discussing the trilemma facing the eigenstate-value link in Section 7.

To establish our quantum probabilism theorem, we make use of an easy corollary exploiting Krein and Milman's result that, for suitable spaces, a compact, convex set $A$ is equal to the closed convex hull of its extremal elements $\partial A$.\footnote{This holds for Hausdorff, locally convex topological vector spaces, and so for any $[0,1]^L$ embedded in $\mathbb{R}^L$ with the product topology. Since $[0,1]^L$ is compact, any closed subset of it is compact in $\mathbb{R}^L$.}
\begin{figure}
    \centering
    \begin{tikzpicture}
        \pgfmathsetmacro{\factor}{4};
        \pgfmathsetmacro{\inter}{1.5};
        \pgfmathsetmacro{\bigc}{.42};
        
        \coordinate [label=below left:0] (a1) at (0*\factor,0*\factor);
        \coordinate [label=above left:1] (b1) at (0*\factor,1*\factor);
        \coordinate [label=below right:1] (a0) at (1*\factor,0*\factor);
        \coordinate [label=right:] (b0) at (1*\factor,1*\factor);
        \coordinate [label=below:$\phi$] (phi) at (.5*\factor,0*\factor);
        \coordinate [label=left:$\psi$] (psi) at (0*\factor,.5*\factor);
        
        %% Back side of box
        \draw[name path = backface] (a0)--(b0)--(b1)--(a1)--cycle;
        
        \node (p1) at (.2*\factor, .2*\factor) {};
        \node (p2) at (.9*\factor, .2*\factor) {};
        \node (p3) at (.5*\factor, .95*\factor) {};
        \node (p'1) at (.3*\factor, .3*\factor) {};
        \node (p'2) at (.8*\factor, .3*\factor) {};
        \node (p'3) at (.6*\factor, .8*\factor) {};
        \draw [thick, fill=gray!20] plot [smooth cycle,tension=1] coordinates {(p1) (p2) (p3)};
        \fill [white] plot [smooth cycle,tension=1] coordinates {(p'1) (p'2) (p'3)};
        
        %\filldraw [white] (.4*\factor,.6*\factor) circle (8pt);
        %\coordinate [label=left:$\textbf{$A$}$] (psi) at (.49*\factor,.7*\factor);
        \coordinate [label=below:$\mathbb{V}$] (psi) at (.5*\factor,.18*\factor);
        
        \coordinate [label=left:$\partial A$] (psi) at (.25*\factor,.7*\factor);
        
    \end{tikzpicture}
    \caption{An illustration of Corollary 1.1 for a language $L$ containing two propositions, $\phi$ and $\psi$.}
    \label{fig:kreincor}
\end{figure}
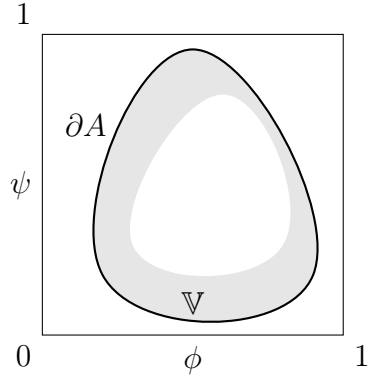
The extremal elements, $\partial A$, are just those elements of $A$ that cannot be expressed as a finite convex sum of more than one element of $A$. Thus:
\begin{quote}
\textbf{Corollary 1.1.} [Proposition 3.1 in \cite{Kuhr2007}] For $\mathbb{V}$ pointwise-closed in $[0,1]^{L}$, if $A\subseteq [0,1]^{L}$ is pointwise-closed and convex, $\mathbb{V}\subseteq A$ and $\partial A \subseteq \mathbb{V} $, then $B\in[0,1]^L$ avoids Dutch books if and only if $B\in A$.
\end{quote}
\noindent
This corollary is illustrated for $L$ containing two propositions in Figure \ref{fig:kreincor}, where $A$ is the oval and its interior, $\partial A$ is the oval, and $\mathbb{V}$ is the oval and the shaded region. In Section 4, we take $A$ to be the set of Born-rule probabilities evaluated over the propositions in a Hilbert-lattice logic and $\mathbb{V}=\partial A$ to be the set of vector states evaluated over these propositions (these concepts will be detailed in Section 3).

We are also concerned to address beliefs which may not live in $\overline{\mathrm{co}}(\mathbb{V})$, but are possibly countably infinite convex sums of elements of $\mathbb{V}$. Fortunately, the ``if'' direction of Theorem 1 holds in this case, even when $\mathbb{V}$ is not pointwise-closed.
\begin{quote}\label{cor:noDutch}
\textbf{Corollary 1.2.} A belief function $B\in [0,1]^{L}$ does not admit a Dutch book if $B$ is a convex sum of elements of $\mathbb{V}$.
\end{quote}

Finally, when establishing the trilemma facing the eigenstate-value link in Section 7, we will make use of the following generalization of the ``only if'' direction of Theorem 1.
\begin{quote}\label{cor:forBell}
\textbf{Corollary 1.3.} If there is some finite $\Gamma \subseteq L$ over which $B$ is defined such that $\mathrm{co}(\{V\restriction_{\Gamma} \mid V \in \mathbb{V} \})$ is pointwise-closed and does not contain $B\restriction_{\Gamma}$, then $B$ is Dutch-bookable.
\end{quote}

K\"{u}hr and Mundici apply these results to any logic whose connectives may be expressed as continuous operations on $[0,1]$ and to generalized multi-value algebras. Our quantum probabilism theorem extends these results in a different direction, applying them to Hilbert-lattice logics.

What this application means, however, will of course depend on how one decides to interpret the formal language of such a logic. In order to motivate our two ways of interpreting Hilbert-lattice logics---as describing outcomes of measurements, on the one hand, and as describing properties of objects, on the other---we turn briefly to the mathematical foundations of quantum mechanics.

%%%%%%%%%%%%
%%%%%%%%%%%%
%%%%%%%%%%%%
\section{Operational and realist interpretations}\label{sec:interpretations}
% \vspace{\baselineskip}
% \noindent
% \textbf{3. Operational and Realist Interpretations.}
Quantum mechanics, like its classical predecessor, is a physical theory that describes \emph{states} of a system and the observable quantities or \emph{observables} of this system. But quantum mechanics seems to disturb the tight link in classical Hamiltonian mechanics between states and the values of observables. Schr\"{o}dinger's wave function $\Psi$ describes the state of a quantum system, but it only fixes distributions over the possible values of observables. Born's rule takes the square-integral of the distribution specified by $\Psi$ over some interval of an observable's possible values to specify the probability of finding the system to have a value for the observable in that interval. Associating states with certain preperations of systems, these Born-rule values match the frequencies we observe for measurements on many identically-prepared systems. But exactly what to say about the values of observables before measurement remains a puzzle.

We consider two divergent attitudes one may adopt in deciding what to say. The operational attitude associates states only with preparations. On this attitude, we take the language of our logics to describe measurement outcomes. The realist attitude remains optimistic that we might be able to salvage a tight link between states and observable quantities prior to measurement. On this attitude, we take the language of our logics to describe objects and their properties.\footnote{On our realist approaches, we will consider ontologies where states of preparation determine truth values for propositions about observable quantities prior to measurement. One may well desire a more relaxed realist approach whereon preparation states determine only a distribution over more general ontic states. For such an approach, see Spekkens's ontological models framework (\citeyear{Spekkens2005}).}

A language for these logics naturally arises from the set of projections $\mathcal{P(H)}$ on a separable Hilbert space $\mathcal{H}$ equipped with an inner product $\langle \cdot, \cdot \rangle$.\footnote{The proceeding description of Hilbert-lattice logics is primarily based on R\'{e}dei's account (\citeyear{Redei1998}).} $\mathcal{H}$ is the vector space of square-integrable functions, and so we treat each wave function $\Psi$ as a vector $\eta\in\mathcal{H}$. Each observable is associated with an operator in the set $\mathcal{B(H)}$ of bounded operators on $\mathcal{H}$. There is an adjoint map $*$ on operators such that $\langle \eta , A \eta' \rangle = \langle  A^* \eta , \eta' \rangle$, and we stipulate that $A\in\mathcal{B(H)}$ is an observable if and only if $A=A^*$. The projections $\mathcal{P(H)}$ are just the idempotent observables. For $B(\mathbb{R})$ the set of all Borel subsets of the real line, we associate with each self-adjoint operator $A\in \mathcal{B(H)}$ a unique function $P^A: B(\mathbb{R}) \to \mathcal{P(H)}$ via the spectral theorem. In this formalism, the Born rule takes the inner product $\langle\eta, P^A(a) \eta \rangle$ to yield the probability for observable $A$ to have a value within the Borel set $a$ when the system is in state $\eta\in \mathcal{H}$.

To capture uncertainty about what state describes how a system was prepared, we use a density operator, $\rho$. For some countable sequence $\{\eta_i\}_{i\in I}$ of states in $\mathcal{H}$, $\rho := \sum_{i} p_i |\eta_i\rangle\langle \eta_i|$, where $p_i \geq 0$ for all $i$, $\sum_{i} p_i = 1$, and $|\cdot\rangle\langle \cdot |$ denotes the Hilbert space's outer product. We may take each $p_i$ to correspond to an agent's degree of confidence that the system was prepared in the state $\eta_i$. If the agent knows $\eta_i$ describes how the system was prepared, then $p_i=1$, and we say $\rho$ is pure. An agent may thus always express these beliefs with a function on the projections $B: \mathcal{P(H)} \to [0,1 ] :: \phi \mapsto \mathrm{Tr}(\rho \phi )$. We call such functions \emph{Born-rule probabilities} and denote the set of these probabilities by $\mathcal{B}$. If $\rho$ is a pure state we say the Born-rule probability it induces is a \emph{Born-rule state}. The set of these we write as $\mathbb{V}_{\mathcal{B}}$.

The set of projections is a propositional language, as it is closed under the logical connectives when these are interpreted as the lattice operations of $\mathcal{P(H)}$ viewed as a Hilbert lattice. Consider a countable set of projections $\{\phi_i\}_{i\in I}$ with ranges $\{\mathcal{G}_i\}_{i\in I}$. Let conjunction be the lattice meet $\bigwedge_i \phi _i$, which is the projection onto the intersection $\bigcap_i \mathcal{G}_i$. Let disjunction be the lattice join $\bigvee_i \phi_i$, which is the projection onto the closed linear subspace spanned by the $\mathcal{G}_i$. Finally, let the negation $\neg\phi$ be the projection onto the set of vectors orthogonal to all vectors in the range of $\phi$. We denote entailment by the partial order $\leq_{\mathcal{H}}$, where $\phi \leq_{\mathcal{H}} \psi$ just in case the range of $\phi$ is included in the range of $\psi$, and $\phi\equiv_{\mathcal{H}}\psi$ just in case $\phi \leq_{\mathcal{H}} \psi$ and $\psi \leq_{\mathcal{H}} \phi$. Under this equivalence, $\mathcal{P(H)}$ is closed under arbitrary conjunctions, disjunctions, and negations.

Now we consider what the operationalist and the realist take the propositions in $\mathcal{P(H)}$ to be. Each $\phi \in \mathcal{P(H)}$ induces an equivalence class of pairs of observables and Borel subsets $\{(A_{\alpha},a_{\alpha})\}$ such that $P^{A_{\alpha}}(a_{\alpha})$ are all the same projection. We say an \emph{operational interpretation} of $\mathcal{P(H)}$ takes $\phi$ to be the proposition:
\begin{quote}
After measurement, the values of $\{A_{\alpha}\}$ for the system lie in $\{a_{\alpha}\}$, respectively.
\end{quote}
The sentence $\phi$ makes a claim about the system immediately after the performance of a measurement. Moreover, as we will see in Section 5, we can choose states of affairs such that $\phi$ is meaningful just in case the properties it references were, in fact, measured. Thus, our operationalist need only ascribe to the system those property-values that are associated with the observed outcomes of actual measurements.

By contrast, \emph{realist interpretations} of $\mathcal{P(H)}$ take $\phi$ to make a claim about a system unperturbed by measurement. We say a \emph{classical-realist interpretation} takes $\phi$ to be the proposition:
\begin{quote}
The values of $\{A_{\alpha}\}$ for the system lie in $\{a_{\alpha}\}$, respectively.
\end{quote}
This approach mirrors the usual interpretation of observable quantities in classical mechanics, taking ascriptions of values for observables like position or momentum to literally be ascribing position-values and momentum-values to objects in the system. We say the \emph{dispositional-realist interpretation} takes $\phi$ to be the proposition:
\begin{quote}
The system is disposed to yield a value of $\{A_{\alpha}\}$ in $\{a_{\alpha}\}$ upon measurement of $\{A_{\alpha}\}$, respectively.
\end{quote}
The relevant properties picked out by this proposition are dispositions of objects in the system to have certain values for other properties once the appropriate measurement is made.

%The dispositional-realist takes $\phi$ to make a claim about what happens after measurement, like the operationalist. Unlike the operationalist, however, the dispositional-realist \emph{also} links the values of properties after measurement to certain dispositional properties possessed by objects in the system before measurement.

Our two realist interpretations do not invoke deep metaphysical commitments. There is quite a bit of flexibility regarding how one interprets the properties ascribed by either sort of proposition. We do impose that the properties ascribed by $\phi$ on a dispositional-realist interpretation are dispositional. However, a property ascribed by some $\phi$ on a classical-realist interpretation, such as position, may be taken to be categorical or relational. Moreover, the property ascribed by $P^A(a)$ on either realist interpretation may be taken to be possessed by an individual particle referenced by the observable $A$, or it may be taken to be possessed by the whole system to which this particle belongs. Finally, while we talk quite explicitly of objects and the properties they possess, this talk certainly should not be taken to rule out various bundle-theoretic accounts of objects. The following arguments apply to any and all of these various ways of fleshing out the metaphysical accounts sketched by either of our realist interpretations of a quantum logic.

% We may only wish to take some sublattice of the projection lattice $\mathcal{P(H)}$ to exhaust the propositions in our calculus. Here, we may make use of von Neumann algebras, which are certain special subalgebras of $\mathcal{B(H)}$. $\mathcal{B(H)}$ is the set of bounded operators; these are operators $A$ such that $\sup_{|| \eta || \leq 1} || A \eta || < \infty$. Note that for $A\in\mathcal{B(H)}$ and $\eta\in\mathcal{H}$, $|| A || := \sup_{||\eta|| \leq 1} || A \eta ||$ defines a norm on $\mathcal{B(H)}$ and that $\mathcal{B(H)}$ is an algebra when the product of operators is defined by composition. Moreover, $\mathcal{B(H)}$ is $*$-closed, or closed under taking adjoints. A von Neumann algebra $\mathcal{M}$ is a $*$-closed subalgebra of $\mathcal{B(H)}$ containing the unit operator $\mathbf{1}$ that is closed in both the norm and the strong operator topologies. The latter constraint means that if there exists some sequence $\{A_{\alpha}\}$ of operators in $\mathcal{M}$ for which there is some $A\in\mathcal{B(H)}$ such that $A_n \eta $ converges to $A\eta$ in $\mathcal{H}$ for all $\eta\in \mathcal{H}$, then $A\in\mathcal{M}$ \cite{Redei2007}. $\mathcal{B(H)}$ is a von Neumann algebra, so $\mathcal{P(M)}\subseteq\mathcal{P(H)}$.

To complete the semantics for Hilbert-lattice logics on one of the above realist interpretations, we will define a set of states of affairs $\mathbb{T}$ and use the standard Tarskian condition to define entailment $\leq_{\mathbb{T}}$. Namely, for $\phi,\psi\in \mathcal{P(H)}$, we define
\begin{eqnarray}
\phi \leq_{\mathbb{T}} \psi &\Longleftrightarrow& \forall \ T\in\mathbb{T} \  [T(\phi) = 1 \Rightarrow T(\psi) =1 ] \label{eq:models1}.
\end{eqnarray}
For operational semantics, we consider a variant of Tarskian entailment that tracks all the possible inferences between meaningful propositions:
\begin{eqnarray}
\phi \leq_{\mathbb{T}} \psi &\Longleftrightarrow& \forall \ T\in\mathbb{T} \  [T(\phi) = 1 \text{ and } T(\psi) \text{ defined}\Rightarrow T(\psi) =1 ]\land \notag \\
& &\exists \ T\in\mathbb{T} \  [T(\phi) = T(\psi) = 1 ] \label{eq:models2}.
\end{eqnarray}
On either definition of entailment, we will require truth-value semantics to be sound and complete. That is, $\leq_{\mathbb{T}}$ should align with $\leq_{\mathcal{H}}$. When this holds, the equivalence relations they induce also align. Truth will preserve our existing notions of entailment and sameness of meaning.

In Sections 5 and 6, we apply this semantically-driven approach to the operational and realist interpretations of $\mathcal{P(H)}$, respectively, and assess the Dutch-bookability of beliefs on the resulting logics. But first, we complete the proof of the promised quantum probabilism theorem that applies to beliefs for these logics given either sort of interpretation.

%%%%%%%%%%%%%%%%
%%%%%%%%%%%%%%%%
%%%%%%%%%%%%%%%%
\section{Quantum probabilism in finite dimensions}\label{sec:quantumprob}
% \vspace{\baselineskip}
% \noindent
% \textbf{4. Quantum Probabilism in Finite Dimensions.}
Our quantum probabilism theorem is the following claim: for any quantum logic $\mathcal{P(H)}$ with $\dim(\mathcal{H})<\infty$, if we take the cognitive evaluations to be the Born-rule states $\mathbb{V}_{\mathcal{B}}$, then the Born-rule probabilities $\mathcal{B}$ are all and only the total belief functions avoiding Dutch books. The proof of this theorem relies on the topological properties of states on $\mathcal{B(H)}$, and so this section is more technical than the others; the reader interested in applications may skip to Section 5.

Recall that a state on $\mathcal{B(H)}$ is a normalized, positive linear functional $\omega : \mathcal{B(H)} \to \mathbb{C} $; that is, $\omega(\mathbf{1})=1$ for the identity element $\mathbf{1}\in\mathcal{B(H)}$ and $\omega(A^*A)\geq 0$ for all $A\in\mathcal{B(H)}$. Let $\mathcal{E}$ be the set of all states in the space $\mathcal{B(H)}^*$ of continuous linear functionals on $\mathcal{B(H)}$; by their linearity, these states are finitely additive.\footnote{Here, a state is finitely additive just in case $\sum_{i=1}^n \omega(\phi_i)=\omega \left (\bigvee_{i=1}^n \phi_i\right )$ for mutually orthogonal $\phi_i$.} $\mathcal{E}$ is convex and compact in $\mathcal{B(H)}^*$ with the weak$^*$ topology, the topology of pointwise convergence for linear functionals (\citealt[p. 53]{Bratteli1987a}). We refer to the extremal elements of this set, $\partial \mathcal{E}$, as pure states; when $\mathcal{H}$ is finite-dimensional, the set of pure states is also weakly$^*$ closed (\citealt[p. 203]{Kadison1991}).

To ensure that states behave like probabilities, we would like to consider just those states that are completely additive over sets of mutually orthogonal projections. We say such states are normal. These are just those states that can be expressed as $\mathrm{Tr}(\rho -)$ for some density operator $\rho$ (\citealt[p. 462]{Kadison1983}). Let $\mathcal{N}$ denote the set of all normal states in $\mathcal{B(H)}^*$. When $\dim (\mathcal{H})<\infty$, it is immediate that all states are normal, since finite additivity implies complete additivity.
%We say that a linear functional $f:\mathcal{B(H)}\to\mathbb{C}$ is normal just in case it satisfies the additional continuity condition $\lim \left ( A_{\alpha} \right ) = f (A)$ for any monotonic increasing net $\{A_{\alpha}\}$ in $\mathcal{B(H)}$ with least upper bound $A$ \cite{Redei2007}. A state is normal if and only if it is completely additive \cite[Theorem 7.1.12]{Kadison1983b}.

For any $\mathcal{B(H)}$, $\mathcal{N}$ is convex and $\partial \mathcal{N}$ is the set of vector states, states of the form $\omega_{\eta}: \mathcal{B(H)}\to \mathbb{C} :: A \mapsto \langle \eta,A\eta \rangle $ for $\eta$ a unit vector in $\mathcal{H}$ (\citealt[p. 177]{Alfsen2012}). For $\dim (\mathcal{H})<\infty$, $\partial \mathcal{N}$ coincides with $\partial \mathcal{E}$, and so all pure states may be expressed as $\mathrm{Tr}(\rho - )$ for some pure density operator $\rho$.

Finally, note that the restriction of the domain of states on $\mathcal{B(H)}$ to the set of projections $\mathcal{P(H)}$ is a map into the Tychonoff cube $[0,1]^{\mathcal{P(H)}}$:
\begin{equation}
r: \mathcal{E} \to [0,1]^{\mathcal{P(H)}}  :: \omega \mapsto \omega \restriction_{\mathcal{P(H)}},
\end{equation}
where $\mathcal{E}$ is given the subspace topology generated by the weak$^*$ topology on $\mathcal{B(H)}^*$, and so is compact. Note that $r(\mathcal{N})$ and $r(\partial\mathcal{N})$ are just the Born-rule probabilities $\mathcal{B}$ and Born-rule states $\mathbb{V}_{\mathcal{B}}$, respectively. We use this fact to prove our main theorem.
\begin{quote}\label{thm:quantumprob}
\textbf{Theorem 2.} For $\dim (\mathcal{H}) < \infty$, a total belief function $B$ on a quantum logic $\mathcal{P(H)}$ with cognitive evaluations $\mathbb{V}_{\mathcal{B}}$ avoids Dutch books if and only if $B\in \mathcal{B}$.
\end{quote}
\noindent
The proof, included in the appendix, proceeds by showing that $r$ is an injective, convexity-preserving closed map, and so $r (\mathcal{E}  )$ is closed and convex, $r(\partial \mathcal{E} )$ is closed, and $r(\partial\mathcal{E})=\partial r(\mathcal{E})$. The result then directly follows from Corollary 1.1.

Furthermore, since every normal state is a convex sum of vector states, it is immediate from Corollary 1.2 that Born-rule probabilities always avoid Dutch books for cognitive evaluations $\mathbb{V}_{\mathcal{B}}$.
\begin{quote}\label{cor:quantumprob}
\textbf{Corollary 2.1.} A belief function $B$ on a quantum logic $\mathcal{P(H)}$ with cognitive evaluations $\mathbb{V}_{\mathcal{B}}$ avoids Dutch books if $B\in \mathcal{B}$.
\end{quote}
\noindent
So long as vector states capture ideal beliefs, then, the Born-rule beliefs avoid Dutch books even when complete and finite additivity do not align.

These results may be applied directly to operational and realist interpretations of the Hilbert-lattice logic $\mathcal{P(H)}$. We consider an operational interpretation first, recovering an analogue of Randall and Foulis's result.

%%%%%%%%%%%%%%%%
%%%%%%%%%%%%%%%%
%%%%%%%%%%%%%%%%
\section{Operational quantum logics on semantic approach}\label{sec:op}
% \vspace{\baselineskip}
% \noindent
% \textbf{5. Operational Quantum Logics on Semantic Approach.}
It is natural for an operationalist to suggest that a suitable bookie must always have the potential to be an actual person. The bookie, then, should determine the ideal beliefs $V_T\in\mathbb{V}$ from the outcomes of a measured system. A human bookie in such a situation will never be able to assign a cognitive evaluation $V_T$ that is defined for propositions $\phi$ and $\psi$ that reference incompatible observables $A$ and $B$, because it is impossible for these observables to be measured simultaneously.

Thus, one option for the operationalist is to take each function $T\in \mathbb{T}$ and each function $V\in\mathbb{V}$ to be defined only for some maximal set $\mathcal{A}$ of commuting projections in $\mathcal{P(H)}$. Consider, for instance, the following toy example of an operational proposition: ``After measurement, the electron has spin up in the $z$-direction.'' We take this proposition to be meaningful, and so to have a truth value, just in case the electron's spin in the $z$-direction was measured. Each $\mathcal{A}$ represents a maximal set of compatible measurements. Thus, each $T\in\mathbb{T}$ will be defined for some maximal set of meaningful sentences.
\begin{quote}
\emph{Operational outcomes.} Every yes-no question in some maximal set of commuting projections has a definite answer. Let $\mathbb{T}_O$ be the set of all functions
$$
T_{\eta,\mathcal{A}}(\phi) =
\begin{cases}
0 &\text{if } \langle \eta, \phi \eta \rangle = 0, \ \phi \in\mathcal{A}  \\
1 &\text{if } \langle \eta, \phi \eta \rangle = 1, \ \phi \in\mathcal{A}  \\
\text{undefined} &\text{otherwise }
\end{cases}
$$
for $\phi \in\mathcal{P(H)}$, $\mathcal{A}\subset \mathcal{P(H)}$ the maximal set of commuting projections generated by $\{|\eta_i\rangle \langle\eta_i|\}$ for $\{\eta_i\}$ an orthonormal basis, and $\eta \in \{\eta_i\}$.
\end{quote}
If we let $\mathbb{T}_O$ specify the entailment relation $\leq_{O}$ for our language $L$ by the modified Tarskian condition (\ref{eq:models2}), then it is not difficult to show that $\phi \leq_{O}\psi$ if and only if $\phi \leq_{\mathcal{H}}\psi$, as desired.

As we noted in Section 2, allowing propositions that are neither true nor false creates two interpretive issues, one for the set of cognitive evaluations and one for the agent's belief-forming strategy. Rather than trying to resolve the former issue definitively, we identify three plausible constraints on ideal beliefs. First, an agent should not hold differing degrees of belief in propositions assigned the same truth value. Second, she should have no degree of belief in meaningless propositions. Third, she should fully believe in true propositions and have no confidence in false ones. We gather these constraints under the heading of \emph{weak truth-fealty}.
\begin{quote}
\emph{Weak truth-fealty.} For all $T\in\mathbb{T}$, the following are satisfied.
\begin{enumerate}
\item If $T(\phi)=T(\psi)$ then either $V_T(\phi)=V_T(\psi)$ or both are undefined.
\item If $T(\phi)$ is undefined then $V_T(\phi)$ is undefined.
\item If $T(\phi)=1$ then $V_T(\phi)=1$ and if $T(\phi)=0$ then $V_T(\phi)=0$.
\end{enumerate}
\end{quote}
Recall $V_T(\phi)$ is undefined when the agent should ideally withhold judgment about $\phi$. Given weak truth-fealty and states of affairs $\mathbb{T}_O$, the ideal beliefs $\mathbb{V}_O$ are uniquely fixed: $\mathbb{V}_O := \{V_{T} \mid T\in\mathbb{T}_O\}$, where $V_T = T.$

We also must address how an agent should form degrees of confidence in propositions she suspects are neither true nor false. Given the propositions in $\mathcal{P(H)}$ on operational interpretation concern only observed outcomes, we naturally suggest that an agent use the empirically-verified Born rule to fix her beliefs. Explicitly, the agent should choose a suitable belief function $B\in\mathcal{B}$ given her degrees of confidence regarding which state describes how the system was prepared. We will subsequently refer to this belief-forming strategy as \emph{Born-fixing}.

As Randall and Foulis have noted at greater length, given Born-fixing, operational quantum logics should avoid Dutch books. These authors use test spaces, a set-theoretic setting for operational probabilities, to establish necessary and sufficient conditions for satisfying a no-Dutch-book condition defined in terms of betting pools (\citeyear[p. 193]{Randall1976}). However, we may quickly review the main result of interest for our purposes---that, on the above operational semantics, all Born-rule probabilities avoid synchronic Dutch books---using our current setup.
\begin{quote}\label{prop:op}
\textbf{Proposition 3.} A belief function $B \in [0,1]^{\mathcal{P(H)}}$ given cognitive evaluations $\mathbb{V}_O$ avoids Dutch books if $B\in\mathcal{B}$.
\end{quote}
\noindent
This result is immediate from Corollary 2.1, as for each testable $\Gamma\subset \mathcal{P(H)}$ given cognitive evaluations $\mathbb{V}_O$, the set of $V\restriction_{\Gamma}$ for $V\in\mathbb{V}_O $ defined over $\Gamma$ is is simply a subset of the set of $V\restriction_{\Gamma}$ for $V\in\mathbb{V}_{\mathcal{B}}$.

As discussed in the introduction, we now abandon the idea that a bookie always has the potential to be some human who can check outcomes. Instead, we imagine our bookie as some hypothetical entity that has perfect knowledge of the state of a quantum system before measurement. We may now use the Dutch book argument as one means of assessing the plausibility of our two candidate ontologies of quantum objects: the eigenstate-value link and the vague-properties approach.

%%%%%%%%
%%%%%%%%
\section{Realist approaches and the coherence of vague properties}\label{sec:real}
% \vspace{\baselineskip}
% \noindent
% \textbf{6. Realist Approaches and the Coherence of Vague Properties.}
In order to apply Dutch books to these ontologies, we detail ways the proponent of the eigenstate-value link and the vague-property theorist may describe states of affairs $\mathbb{T}$ and cognitive evaluations $\mathbb{V}$ for their Hilbert-lattice logics. Each of these may, in principle, be applied to both the classical-realist and dispositional-realist interpretations of the propositional calculus $\mathcal{P(H)}$. However, the eigenstate-value link is traditionally stated as a claim about classical-realist property-ascriptions, and we will default to this interpretation when discussing it.

Loosely, the link says that a system has a sharp or definite value for an observable if and only if it is in an eigenstate of that observable. A state $\eta$ is said to be in an eigenstate of an observable $A$ with eigenvalue $\lambda\in\mathbb{C}$ if $A\eta= \lambda\eta$. On a first pass, then, the link asserts that given a system in the state $\eta$, the proposition ``the value of $A$ for the system is $\lambda$'' is true just in case $A\eta = \lambda\eta$, false just in case $A\eta = \lambda'\eta$ for some $\lambda'\neq\lambda$, and otherwise indeterminate.

There is a small technical hurdle. We have assumed that $\mathcal{H}$ is separable and we have used the spectrum of an observable to represent its possible values. On this approach, observables like position or momentum have no eigenvalues, and so the eigenstate-value rule gives us no guidance for propositions regarding these observables (\citealt[p. 12]{Redei1998}). Thus, we apply the eigenstate-value link just to projections, the propositions in our logics. If a system is in the state $\eta$, the proposition ``the value of $A$ for the system lies in $a$'' has a definite answer just in case $\langle\eta, P^A(a)\eta \rangle$ is zero or one.

This statement of the link does not provide much guidance on how to treat $\phi$ when $\langle\eta, \phi \eta \rangle$ is neither zero nor one. We identify two approaches the defender of the link might have in mind. First, one might suggest in such cases that the proposition $\phi$ is ill-formed, that it has no meaning. One may plausibly ascribe this attitude to Dirac, who suggests the link yields a ``restricted meaning for an observable `having a value'{''} (\citeyear[p. 47]{Dirac1958}).\footnote{\cite{Gilton2016} gives an overview of approaches to the eigenstate-value link espoused by Dirac, Bohm, and von Neumann, among others.} 
% \begin{quote}
% The expression that an observable `has a particular value' for a particular state is permissible in quantum mechanics in the special case when a measurement of the observable is certain to lead to the particular value, so that the state is an eigenstate of the observable. (\citeyear[46]{Dirac1958})
% \end{quote}
We might model this attitude with the following states of affairs.
\begin{quote}
\emph{Eigenstate-value properties (restricted).} Many yes-no questions lack meaning. For $\phi \in \mathcal{P(H)}$, let
$$
\mathbb{T}_{R} = \{T_{\eta} \mid \eta\in\mathcal{H}, ||\eta|| = 1  \}, \qquad T_{\eta}(\phi) =
\begin{cases}
0 &\text{if } \langle \eta, \phi  \eta\rangle = 0\\
1 &\text{if }  \langle \eta, \phi  \eta\rangle = 1 \\
\text{undefined} &\text{otherwise}.
\end{cases}
$$
\end{quote}
Using the standard Tarskian condition (\ref{eq:models1}) to define $\leq_R$, we see straightaway that $\phi \leq_R \psi$ if and only if $\phi \leq_{\mathcal{H}} \psi$. Weak truth-fealty uniquely fixes the cognitive evaluations $\mathbb{V}_R$.

% $\mathbb{T}_O$ and cognitive evaluations $\mathbb{V}_O$ to a classical-realist interpretation of $\mathcal{P(H)}$, whereon the truth of $\phi$ is undefined if $\langle\eta, \phi \eta \rangle$ is neither zero nor one.

But this approach seems unpalatable. We say that cognitive evaluations \emph{restrict expression} when they suggest an agent should not form beliefs about $\phi$ when $\langle\eta, \phi \eta \rangle$ is neither zero nor one. Given such restriction, agents cannot have any degree of confidence whatsoever in many expressions regarding the quantum world. The cognitive evaluations in $\mathbb{V}_R$ clearly restrict expression.

More drastically, this approach implies that propositions ascribing properties to the system gain or lose meaning depending on the state in which the system was prepared. Such contingency of meaning is coherent, perhaps, but it strains intuitions regarding property-ascribing propositions. A change in the weather, one would think, may affect the truth of my assertion that it is raining outside---but surely it cannot make this assertion unintelligible.

When applying states of affairs $\mathbb{T}_O$ to an operationalist interpretation of $\mathcal{P(H)}$, we had a principled reason for both restriction of expression and contingency of meaning. In line with the operationalist's quietist attitude, we required that property-values only be ascribed based on outcomes of performed measurements. For realist interpretations, we venture to treat property-ascriptions quite independently of what measurements are made, and so it is not clear this principle can still be used to cogently justify these constraints.

Thus, we turn to a milder interpretation of the eigenstate-value link. On this second approach, we suggest that if $\langle\eta, \phi\eta \rangle$ is neither zero nor one, then the proposition $\phi$ is meaningful, but it is sent to a truth value that is neither true nor false. This option is modeled by the states of affairs $\mathbb{T}_E$ defined below. We will turn to the question of what an agent's belief-forming strategy for this semantics should be in Section 7. Presently, we note that, assuming weak truth-fealty, agents must either have some constant degree of belief $c\in[0,1]$ in indeterminate propositions or must withhold belief in such propositions, modeled by letting $c$ be undefined.
\begin{quote}
\emph{Eigenstate-value properties (unrestricted).} Many yes-no questions have the answer ``indeterminate.'' For $\phi \in \mathcal{P(H)}$, let
$$
\mathbb{T}_E = \{T_{\eta} \mid \eta\in\mathcal{H}, ||\eta|| = 1  \}, \qquad T_{\eta}(\phi) =
\begin{cases}
0 &\text{if } \langle \eta, \phi  \eta\rangle = 0\\
1 &\text{if }  \langle \eta, \phi  \eta\rangle = 1 \\
2 &\text{otherwise}
\end{cases}
$$
and let
$$
\mathbb{V}_E :=\{V_T \mid T\in \mathbb{T}_E \}, \qquad V_T(\phi) =
\begin{cases}
0 &\text{if } T(\phi)=0\\
1 &\text{if } T(\phi)=1\\
c &\text{if } T(\phi)=2
\end{cases}
$$
for either $c\in [0,1]$ or $c$ undefined.
\end{quote}
We again use (\ref{eq:models1}) to define $\leq_E$. For $\phi,\psi \in \mathcal{P(H)}$, $\phi \leq_{E} \psi$ if and only if $\phi \leq_{\mathcal{H}} \psi$, as desired. This position is presumably what Wallace has in mind in his recent explication of the eigenstate-value link, given that he suggests it ascribes ``completely indefinite positions'' to ``all realistic quantum systems'' (\citeyear[p. 4580]{Wallace2012}).

We need not assign all indeterminate propositions the same truth value, however. A more fine-grained description of indeterminate propositions is afforded by vague properties.
\begin{quote}
\emph{Vague properties}. Many yes-no questions have vague answers. Let 
$$
\mathbb{T}_V = \{T_{\eta} \mid \eta\in\mathcal{H}, ||\eta|| = 1  \}, \qquad T_{\eta}(\phi) = \langle\eta ,\phi\eta \rangle
$$
for $\phi \in \mathcal{P(H)}$.
\end{quote}
Again assuming (\ref{eq:models1}), for all $\phi,\psi \in \mathcal{P(H)}$, $\phi \leq_{V} \psi$ if and only if $\phi \leq_{\mathcal{H}} \psi$.

The vague properties approach takes the Born-rule states to stipulate degrees of truth of realist propositions such as ``the electron has an $x$-position value in the Borel set $a$'' or ``the electron gives rise to an $x$-position value in the Borel set $a$ when its $x$-position is measured.'' Such propositions are assigned a degree of truth equal to their Born-rule value given the state in which the system was prepared. Loose talk of an electron being a ``cloud,'' or having a ``smear'' of position values, is here given a precise formulation as the application of states of affairs $\mathbb{T}_V$ to a classical-realist interpretation of $\mathcal{P(H)}$. Just as we may choose to say that ``the man is bald'' is only somewhat true if the man in question is balding but has tufts of hair around his ears, we say that it is only somewhat true that the electron is located at any given position at a given moment. More extensive treatments of this approach have been provided by, among others, \cite{Lewis2016} and \cite{Pykacz2015}.

Given such states of affairs, there is a ready strategy for how agents should form degrees of belief in the vague propositions in $\mathcal{P(H)}$. We simply suppose that an agent should be precisely as confident in a proposition as the degree to which she expects it is true; this expectation may be expressed by a suitable convex sum of truth valuations.\footnote{\cite{Smith2014} offers a defense of this proposal.} We call this strategy \emph{truth-fixing}. On vague-property semantics $\mathbb{T}_V$, truth-fixing aligns with Born-fixing. In either case, the agent's beliefs should be an appropriate convex sum $B\in\mathcal{B}$ of Born-rule states.

Ideally, the agent should expect the truth value to be the actual one. We capture this desire with a condition of \emph{strong truth-fealty} on cognitive evaluations.
\begin{quote}
\emph{Strong truth-fealty.} For all $T\in\mathbb{T}$: $T=V_T$.
\end{quote}
Given strong truth-fealty, $\mathbb{V}_V =\mathbb{V}_{\mathcal{B}}$. Thus, it is an immediate consequence of Theorem 2 and Corollary 2.1 that, given truth-fixing, all beliefs an agent may have about quantum objects with vague properties avoid Dutch books, and all and only these beliefs are the total ones that avoid Dutch books when $\mathcal{H}$ is finite-dimensional. Note each $T\in\mathbb{T}_V$ does not depend on what measurements are or are not performed; thus, vague property states of affairs applied to classical-realist and dispositional-realist interpretations of $\mathcal{P(H)}$ provide two simple and coherent noncontextual ontologies for quantum mechanics. 

Moreover, vague-property semantics given the dispositional-realist interpretation avoids the measurement problem---there is no question as to why measurements yield particular sharp outcomes. The partial truth of propositions like ``the electron gives rise to an $x$-position value in the Borel set $a$ when its $x$-position is measured'' may be straightforwardly interpreted as the probability the electron \emph{does} in fact yield such a $x$-position-value when its $x$-position is measured. This strategy exploits the standard interpretation of degrees of truth in \L{}ukasiewicz logic, viewing $\mathcal{P(H)}$ as a similar sort of probabilistic logic (\citealt{Pykacz2015}). Of course, this approach is silent about how to conceive of classical properties like position and momentum prior to measurement.

Given its treatment here, the vague properties approach to a classical-realist interpretation of $\mathcal{P(H)}$ has the opposite problem: it is silent about what happens during measurement. One of the virtues of the approach on this front is its flexibility. Prima facie, the ontology offered by the vague properties approach should be easily accommodated by both relative-state and wavefunction-collapse approaches to measurement. We leave the development of this narrative for future work.

%%%%%%%%%%%%%%
%%%%%%%%%%%%%%
\section{Betting with the eigenstate-value link: a trilemma}\label{sec:trilemma}
\begin{figure}
\begin{center}
\begin{tikzcd}
	\text{ } & \Centerstack[c]{weak truth-fealty \\ ($\mathbb{V}_R$ or $\mathbb{V}_E$)} \arrow[dr, "\mathbb{V}_E"]  \arrow[dl, "\mathbb{V}_E"] \arrow[dd, dashed, "\mathbb{V}_R"]  & \text{ } \\
	 \Centerstack[c]{not Born-fixing} & \text{ } &  \Centerstack[c]{Born-fixing } \arrow[d]  \arrow[dl]  \\
    \text{ } & \Centerstack[c]{no beliefs \\ for some $\phi$ }  & \Centerstack[c]{beliefs for all $\phi$; \\ Dutch books}
\end{tikzcd} 
\end{center}
\caption{A sketch of the trilemma facing the defender of the eigenstate-value link. Assuming states of affairs $\mathbb{T}_R$, the defender follows the dashed path; assuming states of affairs $\mathbb{T}_E$ the defender may follow any of the solid paths.}
\label{fig:trilemma}
\end{figure}
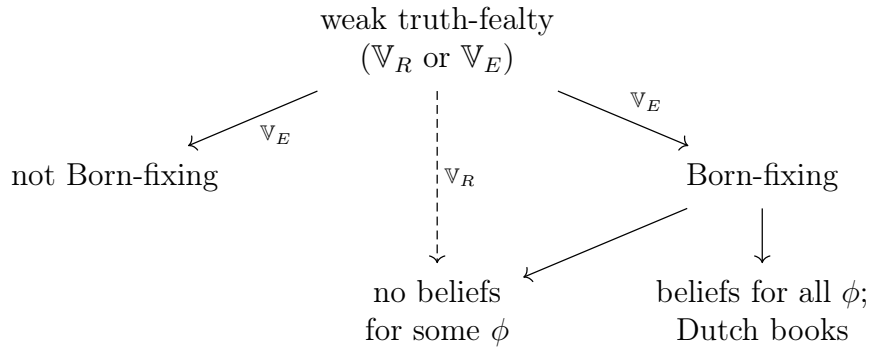

% \vspace{\baselineskip}
% \noindent
% \textbf{7. Betting with the Eigenstate-Value Link: A Trilemma.}
For now, we compare vague-property semantics with our two approaches to eigenstate-value semantics, that with states of affairs $\mathbb{T}_R$ and that with $\mathbb{T}_E$. We have assumed weak truth-fealty to specify the cognitive evaluations $\mathbb{V}_R$ and $\mathbb{V}_E$. Given this mild assumption, the defenders of the eigenstate-value link are faced with a trilemma (Figure \ref{fig:trilemma}). They must choose one of: not using Born's rule to fix their beliefs; using cognitive evaluations that restrict expression; or rendering agents susceptible to Dutch books.

We have already noted that assuming states of affairs $\mathbb{T}_R$ and weak truth-fealty, we are committed to cognitive evaluations $\mathbb{V}_R$, and these cognitive evaluations restrict expression. Thus, the eigenstate-value link defender embracing $\mathbb{T}_R$ is committed to the second horn of the trilemma. The situation is a bit more subtle for $\mathbb{T}_E$.

In Section 6, we postponed the question of how agents should form beliefs given $\mathbb{T}_E$. We address it presently. On the second and third horns of the trilemma, we assume the eigenstate-value defenders recommend Born-fixing. On the first horn, we assume they recommend some other strategy.

The defender may embrace this first horn, suggesting that a mechanism explaining sharp outcomes for measurements may act in such a way that an agent should follow Born-fixing for beliefs about outcomes, but not for beliefs about unseen objects. A minimal modification of truth-fixing affords an option that clearly avoids Dutch books. Supposing $c\in [0,1]$, agents uncertain about the state of a quantum system should choose some appropriate convex sum of functions in $\mathbb{V}_E$; Corollary 2.1 implies that all such belief functions avoid Dutch books.
% Moreover: suppose the eigenstate-value defender recommends that agents form beliefs according to some $B\in \overline{\mathrm{co}}(\mathbb{V}_E)$. Since the range of functions in $\mathbb{V}_E$ is finite, it is immediate that for every $\Gamma\in\mathcal{L}$, $\mathbb{V}_E\restriction_{\Gamma}$ is finite and so pointwise-closed. Thus, by Theorem 2, all and only such $B$ avoid Dutch books.

But it may be desirable to suggest that the agent stick to Born-fixing in this context. After all, one may think that an agent's beliefs about measurement outcomes should directly inform her beliefs about ontic states. For classical Hamiltonian mechanics, this assumption seems nearly tautological. Consider an agent who has a degree of belief $c\in [0,1]$ that if she were to look at a billiard ball at time $t$ she would find it to have value $\lambda$ for its $x$-position. She has the same degree of belief that the ball has $x$-position $\lambda$ at time $t$. If the realist wants to follow this interpretive strategy as closely as possible, then the assumption of Born-fixing is a natural one for her to make. In this light, rejecting Born-fixing merely to avoid Dutch books seems ad hoc.

So suppose we wish to retain the conservatism afforded by the second and third horns. Assuming both weak truth-fealty and Born-fixing, we must choose whether $c$ is in $[0,1]$ or is undefined.

%On the second horn, agents avoid Dutch books, but can have no degree of belief in any indeterminate proposition. On the third horn, the bookie stipulates that agents should have degree of belief $c\in [0,1]$ regarding such propositions, but the agents' beliefs may then be Dutch-bookable.
% \begin{table}
% 	\begin{center}
% 	\begin{tabular}{ l || c | c | c | c | }
% 	    & $a$ & $\neg a$ & $a'$ & $\neg a'$\\
% 	    \hline
% 	  $b$ & 1/2 & 0 & 3/8 & 1/8 \\
% 	  \hline
% 	  $\neg b$ & 0 & 1/2 & 1/8 & 3/8 \\
% 	  \hline
% 	  $b'$ & 3/8 & 1/8 & 1/8 & 3/8 \\
% 	  \hline
% 	  $\neg b'$ & 1/8 & 3/8 & 3/8 & 1/8 \\
% 	  \hline 
% 	\end{tabular}
% 	\end{center}
% \caption{\normalsize\doublespacing Sample probabilities for a Bohm-EPR experiment.}
% \label{tab:BohmEPR}	
% \end{table}

For the second horn, we note that if an agent should have no degree of belief regarding indeterminate propositions, then $\mathbb{V}_E=\mathbb{V}_R$. It is not difficult to use the strategy of the proof of Proposition 3 to show that all beliefs in $\mathcal{B}$ now avoid Dutch books. But it is also clear from our forgoing treatment of $\mathbb{V}_R$ that these cognitive evaluations restrict expression.

For the third horn, we assume $c\in [0,1]$ and consider the spin properties of a pair of entangled electrons. Let $\rho=|\eta\rangle\langle\eta|$ for $\eta=\left (0,1/\sqrt{2},1/\sqrt{2},0\right )$ describe the state of the system on $\mathcal{H}=\mathcal{H}_1 \otimes \mathcal{H}_2$, where each two-dimensional Hilbert space in the tensor product contains states describing just the spin of either particle. Let $X=\{a,a',b,b'\} $ and let $\mathcal{M}=\{\{a,b\},\{a,b'\},\{a',b\},\{a,b'\}\}$; for $C\in\mathcal{M}$, we define
\begin{equation}
P_C := \{x\land y, x\land \neg y, \neg x \land y, \neg x \land \neg y \}_{x,y\in C, x\neq y},
\end{equation}
where, for example, $a\land b = | a_1 \otimes b_2 \rangle \langle a_1 \otimes b_2 | $ given the following eigenspinors in $\mathcal{H}_1$,
\begin{eqnarray}\label{eq:spinors1}
a_1 = \left ( \begin{array}{cc}
         1 \\
         0 \end{array} \right ),
         \quad \neg a_1 = \left ( \begin{array}{cc}
         0 \\
         1 \end{array} \right ),
         \quad a'_1 = \left ( \begin{array}{cc}
         \frac{\sqrt[]{3}}{2} \\
         -\frac{1}{2}\end{array} \right ),
         \quad \neg a'_1 = \left ( \begin{array}{cc}
         \frac{1}{2} \\
         \frac{\sqrt[]{3}}{2} \end{array} \right ),
\end{eqnarray}
and the following eigenspinors in $\mathcal{H}_2$,
\begin{eqnarray}\label{eq:spinors2}
b_2 = \left ( \begin{array}{cc}
         0 \\
         1 \end{array} \right ),
         \quad \neg b_2 = \left ( \begin{array}{cc}
         -1 \\
         0 \end{array} \right ),
         \quad b'_2 = \left ( \begin{array}{cc}
         \frac{1}{2} \\
         \frac{\sqrt[]{3}}{2} \end{array} \right ),
         \quad \neg b'_2 = \left ( \begin{array}{cc}
         -\frac{\sqrt[]{3}}{2} \\
         \frac{1}{2} \end{array} \right ).
\end{eqnarray}
Note that each $P_C$ is a set of four mutually orthogonal projections; we say $P_X$ is the union of these sets. Following Born-fixing, let $B_{\mathrm{Bell}} := \mathrm{Tr}(\rho \phi)$ for $\phi\in\mathcal{P(H)}$; $B_{\mathrm{Bell}}$ is susceptible to Dutch books.
\begin{table}
	\begin{center}
	\begin{tabular}{ l || c | c | c | c  || c | c | c | c |}
	  &$a\land b$ & $a \land \neg b$ & $\neg a \land b$ & $\neg a\land \neg b$ &$a\land b'$ & $a \land \neg b'$ & $\neg a \land b'$ & $\neg a\land \neg b'$\\
	   \hline
	  1&1 & 0 & 0 & 0 & 2 & 2 & 0 & 0\\
      \hline
	  2&0 & 0 & 0 & 1 & 0 & 0 & 2 & 2\\
	  \hline
      3&2 & 0 & 0 & 2 & 2 & 2 & 2 & 2\\
	  \hline
	\end{tabular}
	\end{center}
\caption{\normalsize\doublespacing Some truth valuations compatible with functions in $\mathbb{T}_E$.}
\label{tab:cogval}	
\end{table}
\begin{quote}\label{prop:trad_dutch}
\textbf{Proposition 4.} $B_{\mathrm{Bell}}$ is Dutch-bookable for the cognitive evaluations $\mathbb{V}_E$ given $c\in [0,1]$.
\end{quote}
\noindent
The proof of this proposition notes that Corollary 1.3 implies $B_{\mathrm{Bell}}$ is Dutch-bookable if $B_{\mathrm{Bell}}\restriction_{P_X}$ is not a convex sum of the functions in $\mathbb{V}_{P_X}:=\{ V\restriction_{P_X} \mid V \in \mathbb{V}_E \}$, and proceeds to show that no such convex sum yielding $B_{\mathrm{Bell}}$ can be defined. If the defenders of the eigenstate-value orthodoxy should possess degrees of belief about indeterminate propositions, their beliefs about systems as simple as that of two entangled electrons may be subject to Dutch books.

% In sum: assume the defender of the eigenstate-value link accepts Born-fixing and wants to avoid Dutch books. Regardless of whether this defender embraces states of affairs $\mathbb{T}_O$ or $\mathbb{T}_E$, then, he must accept cognitive evaluations that restrict expression.

Vague properties eschew this trilemma, preserving Born-fixing and avoiding Dutch books while providing a better guide to the quantum world. The proponent of vague properties should always have a degree of belief in a proposition that is neither true nor false, and should thus confidently assert many propositions in which it seems the eigenstate-value defender should have no degree of belief whatsoever.

\section{Conclusions}\label{sec:future}
% \vspace{\baselineskip}
% \noindent
% \textbf{8. Conclusions.}
By extending K\"{u}hr and Mundici's generalized Dutch book theorem to propositional quantum logics $\mathcal{P(H)}$ for $\dim(\mathcal{H})<\infty$, we have supplied a defense of a suitable notion of probabilism for such logics on various interpretations. Our probabilism theorem, like the classical one of de Finetti, does not apply to probabilities for which countable additivity and finite additivity do not coincide. Nonetheless, the classical theorem has been extended by Williamson (among others) to apply to such probabilities by considering infinitary bets which, notably, still only involve a finite amount of money changing hands (\citeyear{Williamson1999}). It is intriguing whether this approach could be extended to arbitrary $\mathcal{P(H)}$.

Still, our present, finitary version of the quantum probabilism theorem unifies Dutch book arguments regarding beliefs about outcomes and beliefs about objects, and it yields significant results for competing accounts of the ontology of quantum mechanics. It renders Born-rule probabilities $\mathcal{B}$ all and only the total beliefs immune to Dutch books on realist interpretations of logics $\mathcal{P(H)}$ ascribing vague properties to quantum objects. In light of the trilemma facing defenders of the eigenstate-value link, vague properties offer a promising alternative.

%%%%%
%%%%%
\section*{Acknowledgements}

Many thanks to Nicholas J. Teh and Brian C. Hall for their invaluable comments on early versions of this work. Thanks also to Pablo Ruiz de Olano and Sebastian Murgueitio Ramirez for the weather example in Section 6. Finally, thanks to all my colleagues in the Notre Dame Philosophy of Physics Research Group and our collaborators at the London School of Economics for their feedback and support.

%\newpage

%%%%%
%%%%%
\section*{Appendix}

Our proofs make use of the following properties of sets of restrictions of the functions in $\mathbb{V}$ to a finite set, which we denote by $\mathbb{V}\restriction_{\Gamma} := \{V\restriction_{\Gamma} \mid V \in \mathbb{V} \}$ for $\Gamma \in \mathcal{L}$, where $\mathcal{L}$ is the family of all finite subsets of $L$ ordered by inclusion.
\begin{quote}\label{prop:finite_res}
\textbf{Claim 1.} If $\mathbb{V}$ is pointwise-closed, then all $\mathbb{V}\restriction_{\Gamma}$ are pointwise-closed.
\begin{proof}
Let $\mathbb{V}$ be a set in the Tychonoff cube $[0,1]^L$ (this space has the topology of pointwise convergence). Pick any $\Gamma\in\mathcal{L}$. All projection functions are continuous, so by the closed map lemma, if $\mathbb{V}$ is closed in $[0,1]^L$, then $\mathbb{V}\restriction_{\Gamma}$ is closed in the Tychonoff cube $[0,1]^{\Gamma}$.
\end{proof}
\textbf{Claim 2.} If $\mathbb{V}\restriction_{\Gamma}$ is pointwise-closed, then its convex hull is pointwise-closed.
\begin{proof}
Pick any $\Gamma=\{\phi_1,\mathellipsis, \phi_n\}$ and suppose $\mathbb{V}\restriction_{\Gamma}$ is pointwise-closed in $[0,1]^{\Gamma}=[0,1]^n$. $\mathbb{V}\restriction_{\Gamma}$ is closed and bounded in $\mathbb{R}^n$, and so it is compact by the Heine-Borel theorem. By Carath\'{e}odory's theorem, $\mathrm{co}(\mathbb{V}\restriction_{\Gamma})$ is closed in $\mathbb{R}^n$ and so is pointwise-closed in $[0,1]^{\Gamma}$.
\end{proof}
\textbf{Claim 3.} For all $\mathbb{V}\restriction_{\Gamma}$ pointwise-closed, $B\restriction_{\Gamma}\in \mathrm{co}(\mathbb{V}\restriction_{\Gamma})$ for all $\Gamma \in \mathcal{L}$ if and only if $B\in\overline{\mathrm{co}}(\mathbb{V})$. 
\begin{proof}
For the ``only if'' direction, note that $B\restriction_{\Gamma} = V^{\Gamma}\restriction_{\Gamma}$ for some $V^{\Gamma}\in \mathrm{co}(\mathbb{V})$. The family of functions $\left \{V^{\Gamma} \right \}_{\Gamma\in\mathcal{L}}$ is a net in $\mathrm{co}\mathbb{V}$ that pointwise converges to $B$, so $B\in\overline{\mathrm{co}}(\mathbb{V})$. For the ``if'' direction, note that if $f$ is in $\overline{\mathrm{co}}(\mathbb{V})$, then some net $\{f_i\}$ in $ \mathrm{co}(\mathbb{V})$ converges pointwise to $f$, and so $\{f_i \restriction_{\Gamma}\}$ is a net in $\mathrm{co}(\mathbb{V}\restriction_{\Gamma})$ converging pointwise to $f\restriction_{\Gamma}$, and so $f\restriction_{\Gamma}\in \overline{\mathrm{co}}(\mathbb{V}\restriction_{\Gamma})$. By each $\mathbb{V}\restriction_{\Gamma}$ closed and Claim 2, each $\mathrm{co}(\mathbb{V}\restriction_{\Gamma})$ is pointwise-closed. Thus: since $B \in \overline{\mathrm{co}}(\mathbb{V})$, $B\restriction_{\Gamma} \in \overline{\mathrm{co}}(\mathbb{V}\restriction_{\Gamma}) = \mathrm{co}(\mathbb{V}\restriction_{\Gamma})$.
\end{proof}
%Since $[0,1]^n$ is itself a closed subspace of $\mathbb{R}^n$ with the product topology, $\mathbb{V}\restriction_{\Gamma}$ is closed in $\mathbb{R}^n$. The product topology is induced by the usual norm on $\mathbb{R}^n$, and so for any $v\in \mathbb{V}\restriction_{\Gamma}$, $0\leq ||v||\leq \sqrt{n}$. Thus, $\mathbb{V}\restriction_{\Gamma}$ is bounded, and so by the Heine-Borel theorem, $\mathbb{V}\restriction_{\Gamma}$ is compact. By Carath\'{e}odory's theorem, the convex hull of a compact set in $\mathbb{R}^n$ is itself compact and so closed. Therefore $\mathrm{co}(\mathbb{V}\restriction_{\Gamma})$ is closed in $\mathbb{R}^n$ and so is pointwise-closed in $[0,1]^{\Gamma}$.
\end{quote}
Now we prove our main results.
\begin{quote}
\begin{proof}[\unskip\nopunct]
\textit{Proof of Theorem 1.} For the ``if'' direction: suppose either that $B$ is some convex sum of elements of $\mathbb{V}$ or that $B\in \overline{\mathrm{co}}\mathbb{V}$. Suppose towards a contradiction that $B$ admits a Dutch book. Then there is some set of sentences $\Gamma=\{\phi_1, \mathellipsis, \phi_n\}$ and some vector of stakes $\langle s_1, \mathellipsis, s_n \rangle \in \mathbb{R}^n$ that satisfies equation (\ref{eq:dutch_gen}). But by $\mathbb{V}$ pointwise-closed and Claims 1 and 3, for some (possibly countably infinite) set of $V_j$, $\sum_j \alpha_j (V_j\restriction_{\Gamma}) = B\restriction_{\Gamma}$ for $\alpha_j\geq 0$, $\sum_j\alpha_j =1$. Thus
\begin{eqnarray*}
& & \sum_j \alpha_j \left (\sum_i s_i \left (V_j (\phi_i)-B(\phi_i) \right )  \right ) < 0 \\
 &\Rightarrow & \sum_i s_i \left ( \sum_j \alpha_j V_j (\phi_i)- \sum_j \alpha_j B(\phi_i) \right )  < 0 \\
&\Rightarrow & \sum_i s_i \left ( B(\phi_i)-B(\phi_i) \right ) < 0.
\end{eqnarray*}
%That these implications hold is trivial for finite sums over $j$; that they hold for countable sums over $j$ follows from the definition of a countable sum in terms of limits and the algebraic limit theorem.
For the ``only if'' direction: suppose that $B \not\in \overline{\mathrm{co}}(\mathbb{V})$. By $\mathbb{V}$ pointwise-closed and Claims 1 and 3, there is some finite set of sentences $\Gamma=\{\phi_1, \mathellipsis, \phi_n\}$ such that $B\restriction_{\Gamma} \not\in \mathrm{co}(\mathbb{V}\restriction_{\Gamma})$. Note that $b=\langle B(\phi_1), \mathellipsis, B(\phi_n) \rangle \in \mathbb{R}^n$ is closed and that, by the proof of Claim 2, $\mathrm{co}(\mathbb{V}\restriction_{\Gamma})$ is a closed convex set in $\mathbb{R}^n$. By the Strong Separating Hyperplane Theorem, there exists a vector $s\in\mathbb{R}^n$ such that $s\cdot v < \alpha$ for all $v\in \mathrm{co}(\mathbb{V}\restriction_{\Gamma})$ and $s\cdot b > \alpha$ for some $\alpha\in \mathbb{R}$ (where $\cdot$ is the inner product for $\mathbb{R}^n$); Equation (\ref{eq:dutch_gen}) is satisfied for this $s$ and $\Gamma$.
\end{proof}
\end{quote}

\begin{quote}
\begin{proof}[\unskip\nopunct]
\textit{Proof of Corollary 1.1.} By the Krein-Milman theorem, $A = \overline{\mathrm{co}} \partial A \subseteq \overline{\mathrm{co}} \mathbb{V} \subseteq \overline{\mathrm{co}} A = A$. Thus $B$ avoids Dutch books if and only if $B\in \overline{\mathrm{co}} \mathbb{V} = A$.
\end{proof}
\end{quote}

\begin{quote}
\begin{proof}[\unskip\nopunct]
\textit{Proof of Corollary 1.2.} Immediate from the proof of the ``if'' direction of Theorem 1.
\end{proof}
\end{quote}

\begin{quote}
\begin{proof}[\unskip\nopunct]
\textit{Proof of Corollary 1.3.} Immediate from the proof of the ``only if'' direction of Theorem 1.
\end{proof}
\end{quote}

\begin{quote}
\begin{proof}[\unskip\nopunct]
\textit{Proof of Theorem 2.} $\mathcal{E}$ is convex and weakly$^*$ closed by Bratteli and Robinson's Theorem 2.3.15 (\citeyear[p. 53]{Bratteli1987a}). Moreover, since $\dim(\mathcal{H}) < \infty$, $\mathcal{E}=\mathcal{N}$, and $\partial \mathcal{E} = \partial \mathcal{N}$ is weakly$^*$ closed by Kadison and Ringrose's Exercise 4.6.67 (\citeyear[p. 203]{Kadison1991}).

We show that $r$ is a continuous closed map. Recall that the weak$^*$ topology on $\mathcal{B(H)}^*$ is the topology of pointwise convergence of linear functionals, and the Tychonoff cube $[0,1]^{\mathcal{P(H)}}$ has the topology of pointwise convergence for functions in the space. So note that if $A$ is closed in $[0,1]^{\mathcal{P(H)}}$ and $\left \{\omega_{\beta} \right \}$ is a net in $r^{-1}(A)$ that converges to $\omega \in \mathcal{E}$, then the net $\left \{\omega_{\beta}\restriction_{\mathcal{P(H)}} \right \} $ converges pointwise to $\omega\restriction_{\mathcal{P(H)}}$  and so converges to $\omega\restriction_{\mathcal{P(H)}}$ in $[0,1]^{\mathcal{P(H)}}$; thus, $\omega \in r^{-1}(A)$. Since $\mathcal{E}$ with the subspace topology is compact and $[0,1]^{\mathcal{P(H)}}$ is Hausdorff, $r$ is a closed map by the closed map lemma. 

By $r$ linear, it preserves convexity. Moreover, $r$ is injective. Note that for every normal state $\omega$ on $\mathcal{B(H)}$, there is some $\rho_{\omega}$ such that $\omega(\phi) = \mathrm{Tr}(\rho_{\omega}\phi)$ and $\rho_{\omega}= \rho_{\omega'}$ if and only if $\omega=\omega'$ by Hall's Theorem 19.9 (\citeyear[p. 424]{Hall2013}). That is, the density operator picked out by a normal state is unique. If two normal states agree over all projections, then they agree over all one-dimensional projections. Thus, note that $\omega$ induces the bounded quadratic form $Q_{\omega}$ on $\mathcal{H}$:
\begin{eqnarray*}
Q_{\omega}:\mathcal{H} \to \mathbb{C} :: \eta \mapsto \omega ( |\eta \rangle\langle \eta |) = \mathrm{Tr}(\rho_{\omega} |\eta \rangle\langle \eta |) = \langle \eta, \rho_{\omega} \eta \rangle.
\end{eqnarray*}
By Hall's Proposition A.63, $Q_{\omega}= Q_{\omega'}$ if and only if $\rho_{\omega}=\rho_{\omega'}$, and so $Q_{\omega}= Q_{\omega'}$  if and only if $\omega=\omega'$ (\citeyear[p. 543]{Hall2013}).

Injective linear functions preserve extremal points. % Suppose $f:X\to Y$ is an injective linear function between topological vector spaces $X$ and $Y$, and let $A$ be convex in $X$. By $f$ linear, $f(A)$ is convex. Consider $y\in f(A)$. By $f$ injective, there is a unique $x\in A$ such that $x=f^{-1}(y)$. Suppose $x \not \in \partial A$. Without loss of generality, $x=\lambda x_1 +(1-\lambda)x_2 $, where $x\neq x_1\neq x_2$ and $0<\lambda<1$. By $f$ linear, $y=\lambda f(x_1) +(1-\lambda)f(x_2)$, and so $y\not\in \partial f(A)$. Now suppose $x\in\partial A $, and suppose towards a contradiction that $f(x)\not\in\partial f(A)$. Then $f(x)=\lambda y_1 + (1-\lambda) y_2$ for $y_1,y_2\in f(A)$, $f(x)\neq y_1 \neq y_2$ and $0<\lambda<1$. But since $f$ is injective, $f^{-1}$ is linear over $f(A)$, and so $x=\lambda x_1+(1-\lambda)x_2$ for some $x_1,x_2\in A$ such that $f^{-1}(y_1)=x_1$ and $f^{-1}(y_2)=x_2$ and $x_1\neq x_2\neq x$. Thus, $x\not\in \partial A$, contradicting our supposition. So $f(\partial A)=\partial f(A)$.
Thus, $r(\partial \mathcal{N})=\partial r( \mathcal{N})$.

It is now immediate that $r(\mathcal{N})=\mathcal{B}$ is closed and convex, and $r(\partial \mathcal{N})=\mathbb{V}_{\mathcal{B}}$ is closed and is equal to the extremal elements $\partial \mathcal{B}$ of $\mathcal{B}$. Thus, Corollary 1.1 completes the proof.
\end{proof}
\end{quote}

\begin{quote}
\begin{proof}[\unskip\nopunct]
\textit{Proof of Corollary 2.1.} By Kadison and Ringrose's Theorem 7.1.12, every normal state is a convex sum of vector states (\citeyear[p. 462]{Kadison1983}). Thus every function in $\mathcal{B}$ is such a convex sum of elements of $\mathbb{V}_{\mathcal{B}}$, and so by Corollary 1.2 every function in $\mathcal{B}$ avoids Dutch books.
\end{proof}
\end{quote}

\begin{quote}
\begin{proof}[\unskip\nopunct]
\textit{Proof of Proposition 3.} For each finite, testable set of propositions $\Gamma \in \mathcal{L}$, the set of $V\restriction_{\Gamma}$ for $V\in\mathbb{V}_O $ defined over $\Gamma$ is simply a subset of $\mathbb{V}_{\mathcal{B}}\restriction_{\Gamma}$. By Corollary 2.1, $B$ does not satisfy equation (\ref{eq:dutch_gen}) for any such $\Gamma$, and so $B$ avoids Dutch books for cognitive evaluations $\mathbb{V}_O$.
\end{proof}
\end{quote}

\begin{quote}
\begin{proof}[\unskip\nopunct]
\textit{Proof of Proposition 4.} Note that $P_X$ contains sixteen propositions and functions in $\mathbb{V}_{P_X}$ range over $\{0,c,1\}$, so $\mathbb{V}_{P_X}$ must be finite, and so it is pointwise-closed in $[0,1]^{P_X}$. Thus by Claim 2, $\mathrm{co}(\mathbb{V}_{P_X})$ is also pointwise-closed, and so by Corollary 1.3, if $B_{\mathrm{Bell}}\restriction_{P_X}$ is not in $\mathrm{co}(\mathbb{V}_{P_X})$, then $B_{\mathrm{Bell}}$ is Dutch-bookable.
    
So suppose towards a contradiction that there is such a convex sum of functions. Note that there are sixteen possible truth valuations on sentences in the set $P_{\{a,b\}}$ that agree with functions in $\mathbb{T}_E$. Table \ref{tab:cogval} gives the only valuations such that $V(a\land \neg b) = V(\neg a\land b)=0$, along with the implied values for $P_{\{a,b'\}}$; these can be deduced from the relevant eigenspinors in equations (\ref{eq:spinors1}), (\ref{eq:spinors2}). Note $B_{\mathrm{Bell}}(a\land \neg b) = B_{\mathrm{Bell}}(\neg a\land b)=0$. So the convex sum yielding $B_{\mathrm{Bell}}\restriction_{P_X}$ must assign nonzero weights only to functions in $\mathbb{V}_{P_X}$ agreeing with a row in Table \ref{tab:cogval}. Say this sum assigns total weights $w_1$ and $w_3$ to the subsets of functions in $\mathbb{V}_{P_X}$ agreeing with row 1 and row 3, respectively. But then $w_1\cdot c + w_3  \cdot c = 3/8=1/8$.
\end{proof}
\end{quote}

%%%%%%%%%%%%%%%
% Bib Options %
%%%%%%%%%%%%%%%
%\newpage

%% BibTeX
\bibliography{contextuality.bib}

\end{document}